\documentclass[aps,twocolumn,superscriptaddress,showpacs,amsmath,amssymb,nofootinbib]{revtex4}
\pdfoutput=1
\usepackage{graphicx,epsf,amsmath}
\usepackage[]{latexsym}
\usepackage{color}

\newcommand{\be}{\begin{eqnarray}}
\newcommand{\ee}{\end{eqnarray}}

\renewcommand{\d}{\mbox{${\rm d}$}} 

\newcommand{\gd}{G_{D}}
\newcommand{\rh}{r_{\rm H}}

\newcommand{\RSD}{R_{D}}

\begin{document}
%
%
\title{Quantum production of black holes at colliders}

\author{Nicusor~Arsene}
\email[]{nicusorarsene@spacescience.ro}
\affiliation{Institute of Space Science, Bucharest,
P.O.~Box MG-23, RO-077125 Bucharest-Magurele, Romania}
\affiliation{Physics Department, University of Bucharest, 
 Bucharest-Magurele, Romania}
\author{Roberto~Casadio}
\email[]{casadio@bo.infn.it}
\affiliation{Dipartimento di Fisica e Astronomia,
Alma Mater Universit\`a di Bologna,
via~Irnerio~46, 40126~Bologna, Italy}
\affiliation{I.N.F.N., Sezione di Bologna, viale Berti~Pichat~6/2, 40127~Bologna, Italy}
\author{Octavian~Micu}
\email[]{octavian.micu@spacescience.ro}
\affiliation{Institute of Space Science, Bucharest,
P.O.~Box MG-23, RO-077125 Bucharest-Magurele, Romania}
\begin{abstract}
We investigate black hole production in $p\,p$ collisions at the Large Hadron Collider
by employing the horizon quantum mechanics for models of gravity with extra spatial dimensions.
This approach can be applied to processes around the fundamental gravitational scale and
naturally yields a suppression below the fundamental gravitational scale and for increasing
number of extra dimensions. 
The results of numerical simulations performed with the black hole event generator BLACKMAX
are here reported in order to illustrate the main differences in the number of expected black
hole events and mass distributions.
\end{abstract}
\maketitle
\section{Introduction}
\label{Sect_Intro}
The possibility to produce black holes (BHs) at particle colliders is directly related to the
question whether the fundamental gravitational scale is somewhere in the few TeV
range, as it was suggested in scenarios with extra spatial dimension, like
the ADD model~\cite{ArkaniHamed:1998rs} and the RS
model~\cite{Randall:1999ee} (see also Ref.~\cite{Maartens:2010ar}
for a comprehensive review).
Above the gravitational scale, it is generally expected that BHs can be created
and finding signatures of their decays would be evidence in favour of these
extra-dimensional models~\cite{cavaglia}. 
During the last years, it was proposed that high energy particle colliders could turn out to be 
huge BH factories~\cite{Giddings:2001bu, Dimopoulos:2001hw}, and
there have actually been many searches to observe the production and
decay of semiclassical and quantum BHs at the LHC~\footnote{A BH is
considered {\em semiclassical\/} if it decays via Hawking radiation, whereas it is generically
called {\em quantum\/} if the decay is not thermal, including the case of a stable
remnant~\cite{remnant}.}.
The ATLAS collaboration looked for events with jet+leptons~\cite{Aad:2014gka,Aad:2013gma}
or dimuon~\cite{Aad:2013lna} in the final state of $pp$ collisions at a centre-of-mass
energy of $\sqrt{s} = 8\,$TeV.
At the same time, the CMS collaboration was searching for energetic multi-particle
final states, as well as for resonances and quantum black holes using the dijet
mass spectra at $\sqrt{s} = 8\,$TeV~\cite{Chatrchyan:2013xva, Khachatryan:2015sja}.
These searches and their results are very important for the community, especially in
the context of the existing extra-dimensional models.
They represent direct comparisons between an experiment and the theoretical
predictions for new physics at these energies and can be used to constrain
the parameters of the models~\cite{gingrich}. 
For example, the CMS collaboration~\cite{Chatrchyan:2013xva}
excluded the production of quantum/semiclassical BHs with masses
below $4.3$ to $6.2\,$TeV (depending on the models) with $95\,\%$ confidence level,
while ATLAS results indicate the threshold mass of the quantum BH to be
larger than $5.3\,$TeV~\cite{Aad:2013gma}.
However, this exclusion limits are strongly dependent on the BH production
cross-section and different decay modes.
\par
Here, we analyse the BH production by employing the modified cross-section
in the ADD model~\cite{ArkaniHamed:1998rs} obtained in Ref.~\cite{Casadio:2015jha}
from the Horizon Quantum Mechanics (HQM) of localised
sources~\cite{Casadio:2013aua,Ctest,RN,CX,ijmpd}.
In fact, this approach was specifically devised to yield the probability that a particle is a BH,
and is therefore perfectly suited to address this issue.
To perform the analysis, we then adapt the BLACKMAX code~\cite{Dai:2007ki}, one of the
most powerful and widely used BH event generators, which includes different scenarios
like tension/tensionless rotating/non-rotating BHs~\footnote{Examples of other available BH
event generators are CHARYBDIS2~\cite{Frost:2009cf}, QBH~\cite{Gingrich:2009da}
and CATFISH~\cite{Cavaglia:2006uk}.}. 
The results of our findings will be presented in Section~\ref{results}.
Before that, we familiarise the reader with the HQM and
provide some useful references for a more in-depth study of the formalism in
Section~\ref{Sect_HWF}.
\section{Horizon Quantum Mechanics}
\label{Sect_HWF}
The HQM for static sources~\cite{Casadio:2013aua,Ctest,RN,CX}
was extended to higher dimensions in Ref.~\cite{Casadio:2015jha,ijmpd},
which can be naturally applied to BHs in the ADD scenario.
Let us start by considering the wave-function for a localised massive particle as given by
a spherically symmetric Gaussian wave-packet of width $\ell$ in $D$ spatial dimensions
[representing a source in a ($D+1$)-dimensional space-time]
\be
\psi_{\rm S}(r)
=
\frac{e^{-\frac{r^2}{2\,\ell^2}}}{(\ell\, \sqrt{\pi})^{D/2}}
\ ,
\label{Gauss}
\ee
whose form in momentum space is 
\be
\tilde{\psi}_{\rm S}(p)
=
\frac{e^{-\frac{p^2}{2\,\Delta^2}}}{(\Delta\, \sqrt{\pi})^{D/2}}
\ ,
\label{momGauss}
\ee
where $\Delta=\hbar/\ell=m_D \,\ell_D/\ell$,
$m_D$ is the fundamental gravitational mass and $\ell_D=\hbar/m_D$
represents the corresponding length scale.
We study the simplest case and assume that, when a BH forms, it will be
described by the ($D+1$)-dimensional Schwarzschild metric
\be
\d s^2
&=&
-\left(1-\frac{\RSD}{r^{D-2}}\right) \, \d t^2
+\left(1-\frac{\RSD}{r^{D-2}}\right)^{-1} \, \d r^2
\nonumber
\\ 
&&
+ r^{D-1} \, \d\Omega_{D-1}
\ ,
\ee
where the classical horizon radius is given by
\be
\RSD(M)
=
\left(\frac{2 \, \gd \, M}{|D-2|}\right)^{\frac{1}{D-2}}
\ ,
\label{SchwD}
\ee
and
$
G_D
=
{\ell_D^{D-2}}/{m_D}
$
represents the fundamental gravitational constant in this ADD scenario. 
\par
We now consider the mass-shell relation in flat space, 
$p^2 = E^2-m^2$, where $m$ is the rest mass of the source,
and express $E$ in terms of the above horizon radius~\eqref{SchwD}, 
$\rh=R_D(E)$.
After using these results in Eq.~\eqref{momGauss}, the
normalised horizon wave-function reads~\cite{Casadio:2015jha}
\be
\psi_{\rm H}(\rh)
=
\mathcal{N}\,
\Theta(\rh-R_D(m))\,
e^{
-\frac{(D-2)^2\,m_D^2}{8\,\Delta^2}\left(\frac{\rh}{\ell_D}\right)^{2(D-2)}}
\!\!\!\!\!\!
\ ,
\label{HWF}
\ee
where the normalisation $\mathcal{N}$ is obtained from the Schr\"odinger scalar product 
in $D$ spatial dimensions and the step function ensures that the gravitational radius
$\rh\ge R_D(m)$, since $m$ is the minimum energy eigenvalue contributing to the packet.
We can now calculate the probability for the particle to be a BH,
\be
P_{\rm BH}
\,=\,
\int_0^\infty
\mathcal{P}_<(r<\rh) \, \d\rh
\ .
\label{PBH}
\ee
where $\mathcal{P}_<(r<\rh)$ represents the probability density for the
particle to be inside its own gravitational radius $\rh$ and is the product of two factors:
the probability for the particle to be located inside a $D$-ball of radius $\rh$
and the probability density that the horizon radius equals $\rh$. 
In this particular case, the BH probability depends on the Gaussian width $\ell$,
particle mass $m$ and number of spatial dimensions $D$.
We can further assume $\ell= \lambda_m=m_D\, \ell_D / m$
is the Compton length of the source, which represents the minimum uncertainty 
in its size, so that $\Delta=m$ and the probability only depends on $m$ and the number
of dimensions $D$~\cite{Casadio:2015jha},
\be
&
P_{\rm BH}(m;D)
=
\strut\displaystyle
\left(\frac{D-2}{2}\right)^{\frac{2}{D-2}}
\frac{(m/m_D)^{\frac{D}{D-2}}}{\Gamma\left(\frac{D/2}{D-2},1\right)
\Gamma\left(\frac{D}{2}\right)}
\qquad
&
\label{PBHexplicit}
\\
&
\times
\!\!\!
{\strut\displaystyle
\int\limits_{R_D(m)}^\infty}
\!\!\!
\gamma\!\left(\frac{D}{2},\frac{m^2\,\rh^2}{m_D^2\,\ell_D^2}\right)
e^{-
\left[\frac{(D-2)\,m_D}{2 \,m}\right]^2
\left(\frac{\rh}{\ell_D}\right)^{2(D-2)}}
\frac{\rh^{D-1}}{\ell_D^D} \, \d\rh
\ ,
&
\notag
\ee
where $\Gamma(a,b)$ is the upper incomplete gamma function and
$\gamma(a,b)$ the lower incomplete gamma function.
The above expression can be computed numerically and is displayed in Fig.~\ref{PBH}
for $D = 5$, $7$ and $9$.
\begin{figure}[t]
\centering
\includegraphics[scale=0.4]{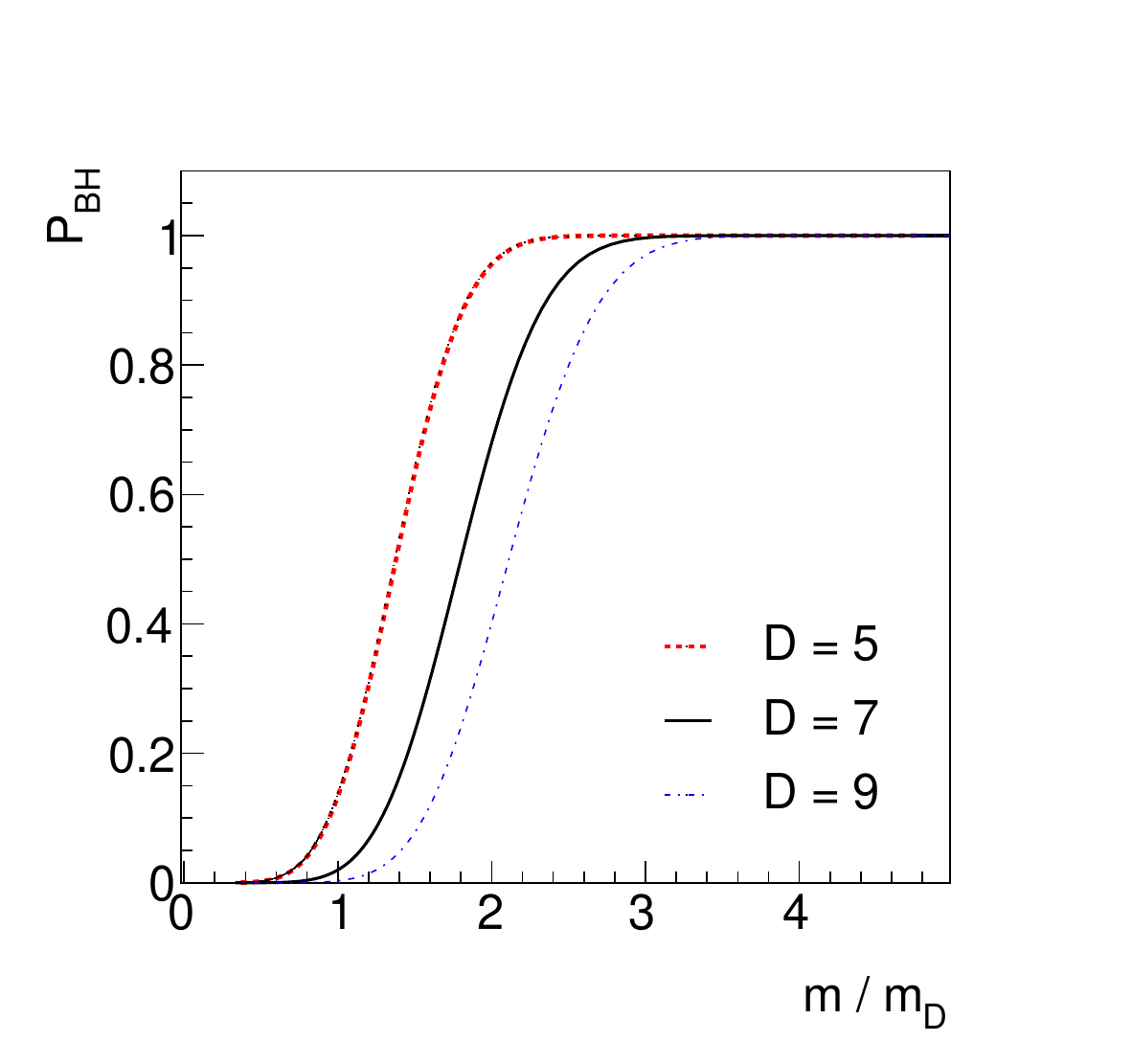}
\caption{Probability~\eqref{PBHexplicit} for a particle described by a Gaussian
wave-function of mass $m$ to be a BH for different
numbers of spatial dimensions $D$.}
\label{PBH}
\end{figure}
\par
There are a few important observations regarding this result.
First of all, like in $D=3$, there is no sharp threshold for BH formation,
but the BH probability drops very fast for $m< m_D$ (or, equivalently,
$\ell>\ell_D$).
Moreover, for any given mass, say $m\simeq m_D$, the probability
$P_{\rm BH}(m;D)$ decreases for increasing values of $D$.
In the next section, we will focus on expressing these differences
in a more quantitative way.
\section{Cross section $p\,p \rightarrow {\rm BH}$}
\label{results}
We will now focus on the implications for BH searches at the LHC. 
As stated in the Introduction, we performed the numerical simulations using
BLACKMAX~2.02.0 and considering tensionless non-rotating BHs.
In the standard configuration, BLACKMAX employs the BH production
cross-section
\be
\sigma_{\rm BH}(E) = b_{D}^2\, \pi\, R_D^2(E)
\ ,
\label{sigmaBH}
\ee
where $b_{D} = 2\,\left[1+(D-1)^2/4\right]^{\frac{1}{2-D}}$
and $R_D$ is the horizon radius~\eqref{SchwD}.
Among other parameters, one can set the values of the
fundamental gravitational scale $m_D$ and the minimum BH mass
$m^{\rm min}$.
In fact, it is important to remark that no threshold
of BH production is fixed in the standard scenarios,
although one expects that BHs do not form below a certain mass
because of quantum fluctuations,
and one can at best constrain $m^{\rm min}$ from the data. 
Typically, we shall consider $m^{\rm min}\gtrsim m_D$
in order to ensure that no BH is produced with a mass below
$m_D$ in the standard case.
\par
\begin{figure*}[ht]
\centering
\includegraphics[scale=0.88]{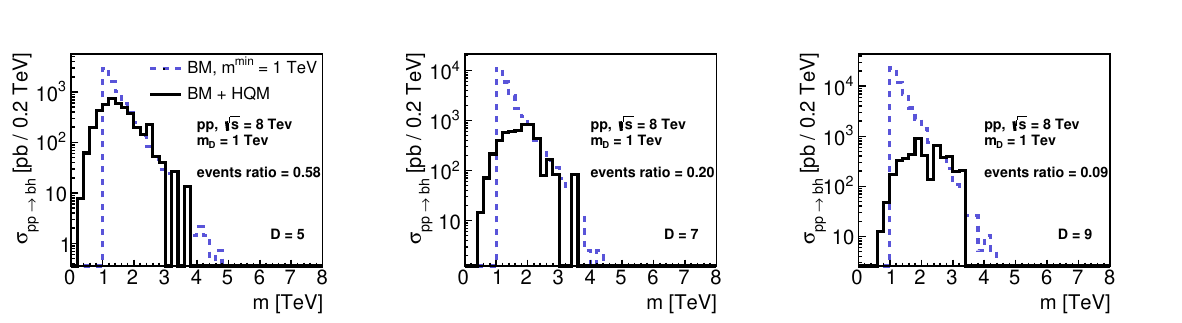}
\caption{BH mass distribution, weighted by the cross-section for $p\,p$ collision 
at  $\sqrt{s} = 8\,$TeV, for a fundamental gravitational mass $m_D = 1\,$TeV
and number of spatial dimensions $D = 5$, $7$ and $9$.
Blue dashed lines are for the standard scenario with cross-section~\eqref{sigmaBH}
and $m^{\rm min} = m_D$;
continuous black lines represent the modified case~\eqref{cross_sect} and
\textit{events ratio\/} gives the corresponding suppression factor.
Each curve is an average over $10^4$ simulations of tensionless
non-rotating BHs in BLACKMAX~2.02.0.}
\label{Mpl1_Mmin1_8TeV_comparison_}
\end{figure*}
\begin{figure*}[ht]
\centering
\includegraphics[scale=0.88]{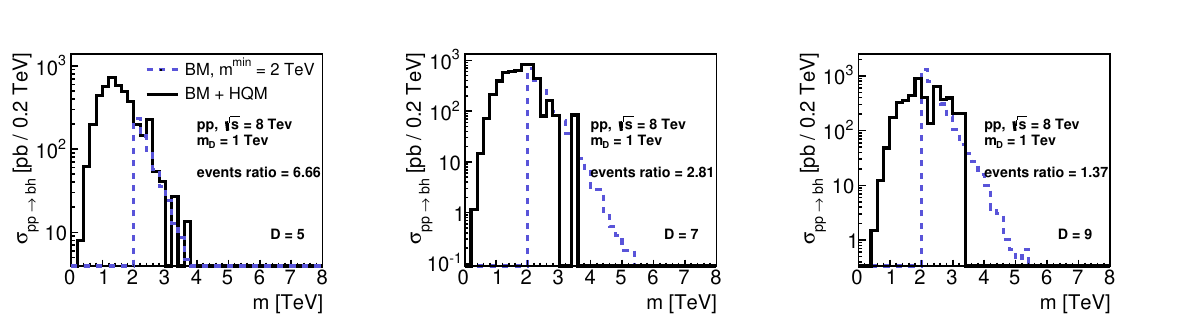}
\caption{BH mass distribution, weighted by the cross-section for $p\,p$ collision 
at  $\sqrt{s} = 8\,$TeV, for a fundamental gravitational mass $m_D = 1\,$TeV
and number of spatial dimensions $D = 5$, $7$ and $9$.
Blue dashed lines are for the standard scenario with cross-section~\eqref{sigmaBH}
and $m^{\rm min} = 2\,m_D$;
continuous black lines represent the modified case~\eqref{cross_sect} and
\textit{events ratio\/} gives the corresponding suppression factor.
Each curve is an average over $10^4$ simulations of tensionless
non-rotating BHs in BLACKMAX~2.02.0.}
\label{Mpl1_Mmin2_8TeV_comparison_}
\end{figure*}
In the HQM picture of Refs.~\cite{Casadio:2015jha,ijmpd},
the effective BH production cross-section is instead given by
\be
\sigma_{\rm HQM}(E)
=
P_{\rm BH}(E)\,\sigma_{\rm BH}(E)
\ ,
\label{cross_sect}
\ee
where $P_{\rm BH}(E)=P_{\rm BH}(m=E;D)$ is the probability~\eqref{PBHexplicit}
for a particle with energy $E$ to be a BH,
while $\sigma_{\rm BH}$ is still given by Eq.~\eqref{sigmaBH}.
Note that there is now no need for imposing a minimum BH mass, since $P_{\rm BH}$
acts as a proper quantum regulator.
In order to implement this improved cross-section, we considered the fundamental
gravitational mass scale to have the same value as in the standard case, and
set a minimum BH mass $m^{\rm min} = 0.2\,m_D$ for computational
convenience~\footnote{We checked the final results do not change significantly
when lowering $m^{\rm min}$ even further.}. 
Finally, we added a subroutine to BLACKMAX which weighs the standard BH mass
distributions with the probability $P_{\rm BH}(E)$.
\par
We first illustrate the typical differences between the simulations that employ the standard
cross-section~\eqref{sigmaBH} and the HQM cross-section~\eqref{cross_sect}
in Figs.~\ref{Mpl1_Mmin1_8TeV_comparison_} and \ref{Mpl1_Mmin2_8TeV_comparison_}.
Later on, we will investigate more general cases and include comparisons to
the current bounds on $m_D$ and $m^{\rm min}$ by the ATLAS and CMS
collaborations.
The blue dashed lines are obtained by employing
the standard BH production cross-section~\eqref{sigmaBH} for $p\,p$ collisions
at $\sqrt{s} = 8\,$TeV, with $m_D = 1\,$TeV,
and setting the minimum BH mass to $m^{\rm min} = m_D$
(in Fig.~\ref{Mpl1_Mmin1_8TeV_comparison_}) or $m^{\rm min} = 2\,m_D$
(in Fig.~\ref{Mpl1_Mmin2_8TeV_comparison_}). 
The continuous black lines in the same plots represent the analogous BH mass distribution
derived from the modified production cross-section~\eqref{cross_sect}. 
First of all, in agreement with Fig.~\ref{PBH} and the HQM approach~\cite{Ctest, RN, Casadio:2015jha},
BHs with masses below the fundamental scale of gravity are now possible and no sharp threshold
effect like the one forced in the standard case exists.
For $m^{\rm min} = m_D= 1\,$TeV, the HQM cross-section~\eqref{cross_sect} leads to a
significant suppression of BH production, whereas for $m^{\rm min} = 2\,m_D= 2\,$TeV
the situation is reversed.
Besides investigating how the differential production cross-section varies with the value of the
resulting BH mass, we can also compare the total cross-sections in the two cases.
This comparison is given by the \textit{events ratio}, the ratio between the HQM
total production cross-section relative to the standard case, which we also display in the plots. 
In particular, the event ratio is smaller than one (thus signalling a suppression)
for $m^{\rm min} = m_D= 1\,$TeV, but larger than one (indicating an enhancement)
for $m^{\rm min} = 2\,m_D= 2\,$TeV. 
We also notice that this ratio is always smaller for larger $D$, again in agreement with
Fig.~\ref{PBH}.
This can be viewed as a check which makes us confident that our numerical simulations
are accurate. 
\par
\begin{figure}
\centering
\includegraphics[scale=0.55]{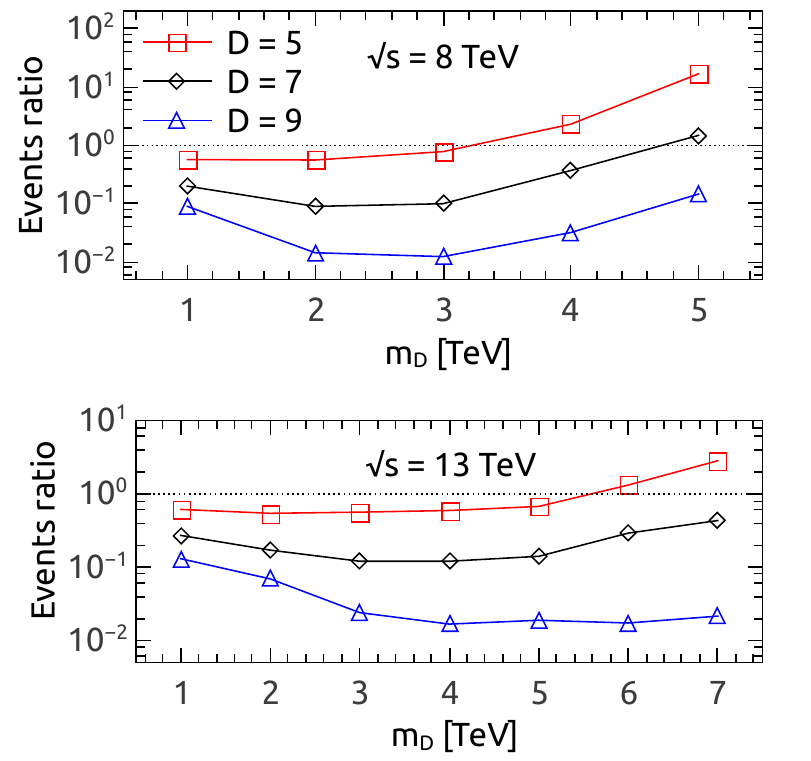}
\caption{Dependence of the \textit{events ratio} on the value of $m_D=m^{\rm min}$
for $\sqrt{s} = 8\,$TeV (top panel) and $\sqrt{s} = 13\,$TeV (bottom panel).
Note the logarithmic scale on the vertical axis.}
\label{summary_8TeV_}
\end{figure}
\begin{figure}
\centering
\includegraphics[scale=0.55]{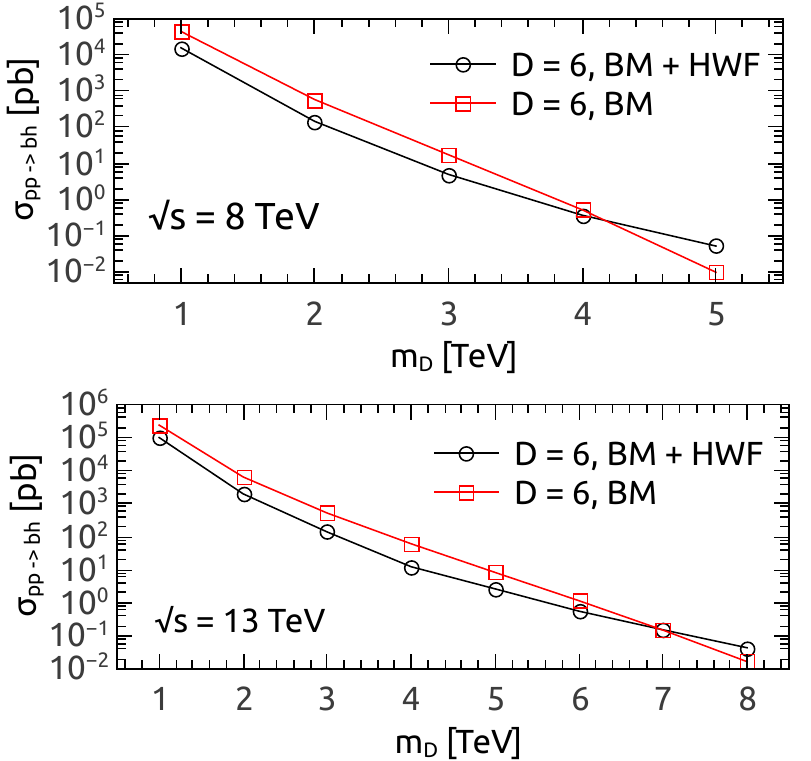}
\caption{Dependence of the total cross-section $\sigma_{p\,p \rightarrow {\rm BH}}$
on $m_D$ for $D = 6$ and $\sqrt{s} = 8\,$TeV (top panel) or $\sqrt{s} = 13\,$TeV
(bottom panel).
Red squares represent the standard cross-section,
while black circles represent the HQM values.}
\label{sigma_Mpl_8TeV_}
\end{figure}
In light of these preliminary results, it appeared interesting to study how the \textit{events ratio\/}
depends on the value of the fundamental gravitational scale for different numbers of spatial dimensions.
This analysis is presented in Fig.~\ref{summary_8TeV_} for $m_D=m^{\rm min}$. 
We see that, in this case, regardless of the value of $m_D$, this ratio is always smaller for larger
number of extra-dimensions (at the same value of the gravity scale).
Another feature we notice is that, for all numbers of spatial dimensions, from around $m_D = 2\,$TeV
for $\sqrt{s} = 8\,$TeV (respectively $m_D = 4\,$TeV for $\sqrt{s} = 13\,$TeV) the \textit{events ratio}
starts to increase with $m_D$, eventually crossing unity from below.
Even though it seems that more BHs are produced in the HQM scenario than in the
standard case for higher values of $m_D$, we have to remember that the total cross-section
decreases with $D$ and the number of expected BH events remains very small.
This dumping of the total cross-sections for increasing $m_D$ is exemplified
in Fig.~\ref{sigma_Mpl_8TeV_} for $D=6$ (but the same behaviour holds in all cases):
for smaller values of $m_D$, the standard production is larger than the HQM expectation,
and the two cross at a relatively large value of $m_D$ (that depends on $D$).
As one can see, even where $\sigma_{p+p \rightarrow {\rm BH}}$ predicted by the HQM
is larger, the actual values of the total cross-sections are very small.
Fig.~\ref{sigma_Mpl_8TeV_} shows, for instance, that when $\sqrt{s} = 8\,$TeV, the cross-sections
are of the order of $10^4\,$pb for $m_D = 1\,$TeV, but reduce to about $1\,$pb for $m_D = 4\,$TeV.
\par
\begin{figure*}[t]
\centering
\includegraphics[scale=0.9]{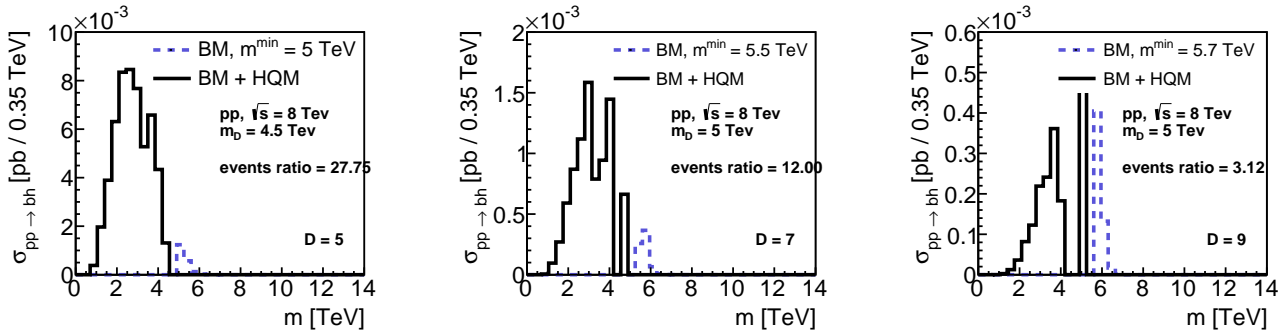}
\caption{BH mass distributions and \textit{events ratio\/} at the current LHC lower bounds for $m_D$
and $m^{\rm min}$.
}
\label{LHC_bounds}
\end{figure*}

So far we mostly analysed cases with $m^{\rm min}=m_D$ in the standard scenario and compared
with the HQM predictions for the same $m_D$.
The tables in Appendix~\ref{events_ratio} present the dependence of the \textit{events ratio}
on $m^{\rm min}$ and $m_D$, taken to be independent parameters for the standard scenario,
for the same $m_D$ in the HQM case.
The thick black lines in the tables show where the \textit{events ratio\/} crosses over one:
below the lines, the HQM predicts less BH events, whereas the standard simulations
predict less such events above the lines. 
It is in particular interesting to compare the number of events predicted by the HQM
with the number of events one expects to see in the standard case for the current lower
bounds imposed by the LHC collaborations on $m_D$ and on the minimum mass that BHs
can have~\cite{Aad:2014gka,Aad:2013gma, Aad:2013lna, Chatrchyan:2013xva, gingrich}.
In Fig.~\ref{LHC_bounds}, we show a comparative analysis between the BH production
cross-sections (which ultimately translate into the number of events expected to be produced
at the LHC), and the distributions of BH masses for these two cases.
In the plots, we assumed the strongest lower bounds on $m_D$ and $m^{\rm min}$
available~\cite{Chatrchyan:2013xva, gingrich}, and as usual compared with the HQM
predictions for the same $m_D$.
It immediately appears that the HQM predicts more BH events.
Upon examining Fig.~\ref{LHC_bounds} further, one also notices that the BH mass
distributions differ: the HQM predicts that most BHs are produced with masses
smaller than the values expected in the standard scenario.
This is no surprise, given that we assumed the same value for $m_D$ in the two
scenarios, the lower bounds imposed by the LHC groups on $m^{\rm min}$
are stronger (higher values) than the bounds imposed on $m_D$, and that BHs with
mass below $m_D$ are possible in the HQM.
\begin{figure*}[t]
\centering
\includegraphics[scale=0.9]{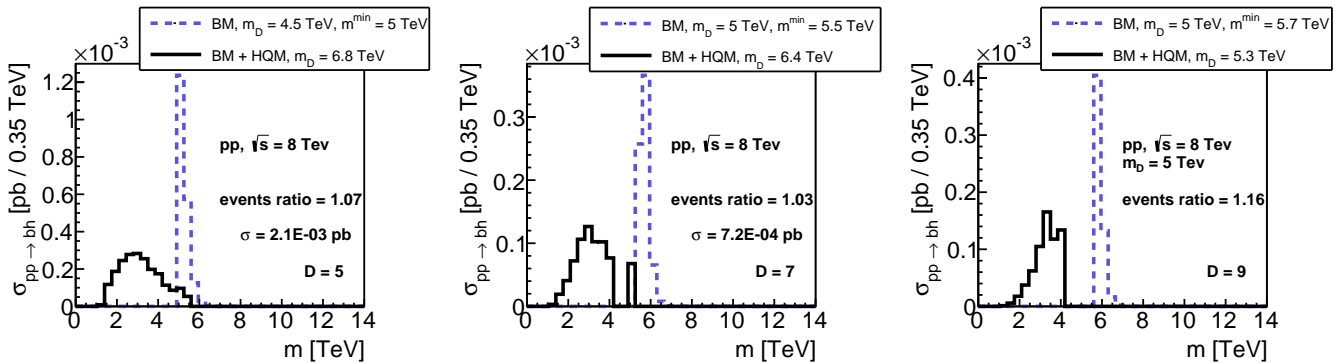}
\caption{HQM lower bound on $m_D$ obtained by requiring \textit{events ratio} be approximately
equal to one.
}
\label{LHC_bounds_v1}
\end{figure*}
\par
Since the HQM yields no minimum BH mass, the above comparison is not completely
significant, and one should actually constrain only $m_D$ using experimental data
in this scenario.
We then determined the value of $m_D$ in the HQM formalism for which the number
of BH events is expected to be the same as in the standard case at the current bounds. 
This means we set the BLACKMAX parameters for the standard case equal to the LHC
bounds, and changed the value of $m_D$ in the HQM simulations.
The results are presented in Fig.~\ref{LHC_bounds_v1}, from which one
can see the \textit{events ratio} is roughly equal to one for $m_5\simeq 6.8\,$TeV,
$m_7\simeq 6.4\,$TeV, $m_9\simeq 5.3\,$TeV.
In all cases, the HQM lower bounds on $m_D$ therefore appear stronger than
in the standard case.  
The number of BH events expected at the LHC is about the same as in the corresponding
standard scenario, but with a very different distribution of masses (most of the BHs are
produced with masses lower than the minimum BH mass in the standard case).
We thus caution our readers that in deriving these HQM lower bounds on $m_D$,
we neglected the impact that the different HQM distributions of BH masses may have on the
likelihood for the LHC collaborations to detect them.
In fact, before the LHC collaborations reached the current limits, they also
scanned the parameter space below those values. 
\par
\begin{figure*}
\centering
\includegraphics[scale=0.9]{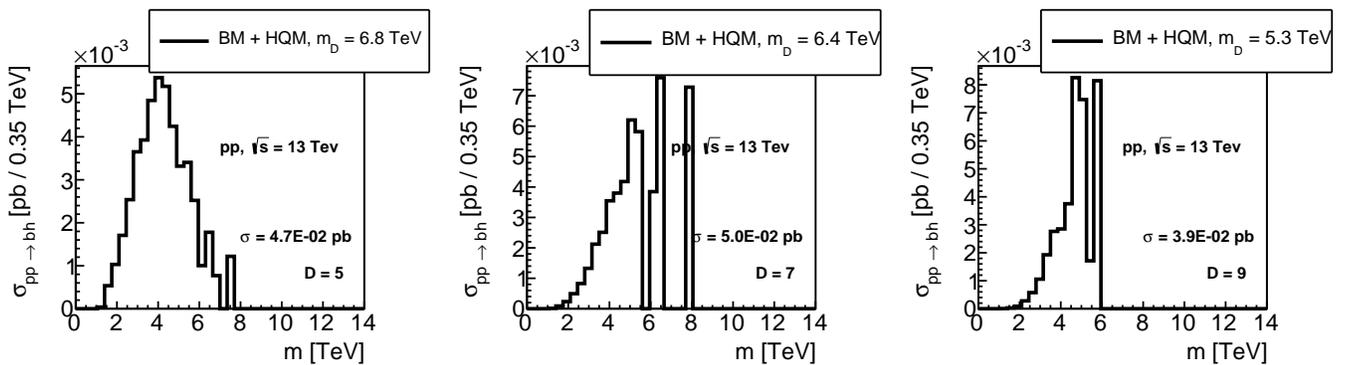}
\caption{HQM estimated BH mass distributions at $\sqrt{s} = 13\,$TeV with
$m_5=6.8\,$TeV, $m_7=6.4\,$TeV and $m_9=5.3\,$TeV.
}
\label{LHC_bounds_13TeV}
\end{figure*}
With that disclaimer in mind, we can still consider the HQM bounds on $m_D$
and simulate the BH production at $\sqrt{s} = 13\,$TeV.
Fig.~\ref{LHC_bounds_13TeV} shows the expected distribution of BH masses in this case,
with cross-sections of the order of a few times $10^{-2}\,$pb. 
\section{Conclusions}
We investigated the implications of the HQM on the BH production cross-sections
at the LHC in the context of the ADD models with extra spatial dimensions.
We used BLACKMAX to perform numerical simulations allowing us to compare
both the production cross-sections and the resulting BH mass distributions.
The \textit{events ratio\/} was used to express quantitatively the differences in the
BH production cross sections in the two cases.
We find that this ratio is always smaller for larger number of 
extra-dimensions, and that in each case it eventually increases with $m_D$ 
until it becomes larger than one.
A wide range of cases are presented in Fig.~\ref{summary_8TeV_}
and the tables in Appendix~\ref{events_ratio}. 
When looking at the distribution of BH masses, in particular, we find that the HQM
predicts most BHs are produced with masses smaller than the values expected 
in the standard scenario.
\par
We also compared the cross section predicted by the HQM with the standard one 
for the current lower bounds imposed by the LHC collaborations on the fundamental 
gravity scale $m_D$ and $m^{min}$.
We found that in the HQM case more BHs are produced.
We thus determined new lower bounds on $m_D$, by finding the value of the 
fundamental gravity scale at which the number of BH events is expected to be the
same as in the standard case at the current bounds: 
$m_5\simeq 6.8\,$TeV, $m_7\simeq 6.4\,$TeV, $m_9\simeq 5.3\,$TeV. 
Finally, we calculated the BH production cross sections for these new lower bounds
at $\sqrt{s} = 13\,$TeV.
It will also be interesting to investigate the implications of the HQM for BH remnants
at the LHC~\cite{Alberghi:2013hca}, or other modified decay channels~\cite{cha1},
but we leave this analysis for future works. 
\subsection*{Acknowledgments}
R.C.~is partly supported by the INFN grant FLAG. N.A. and O.M. were supported by the grant LAPLAS 4. 
\appendix
\section{Events ratio}
\label{events_ratio}
\begin{figure*}
\centering
\includegraphics[scale=0.9]{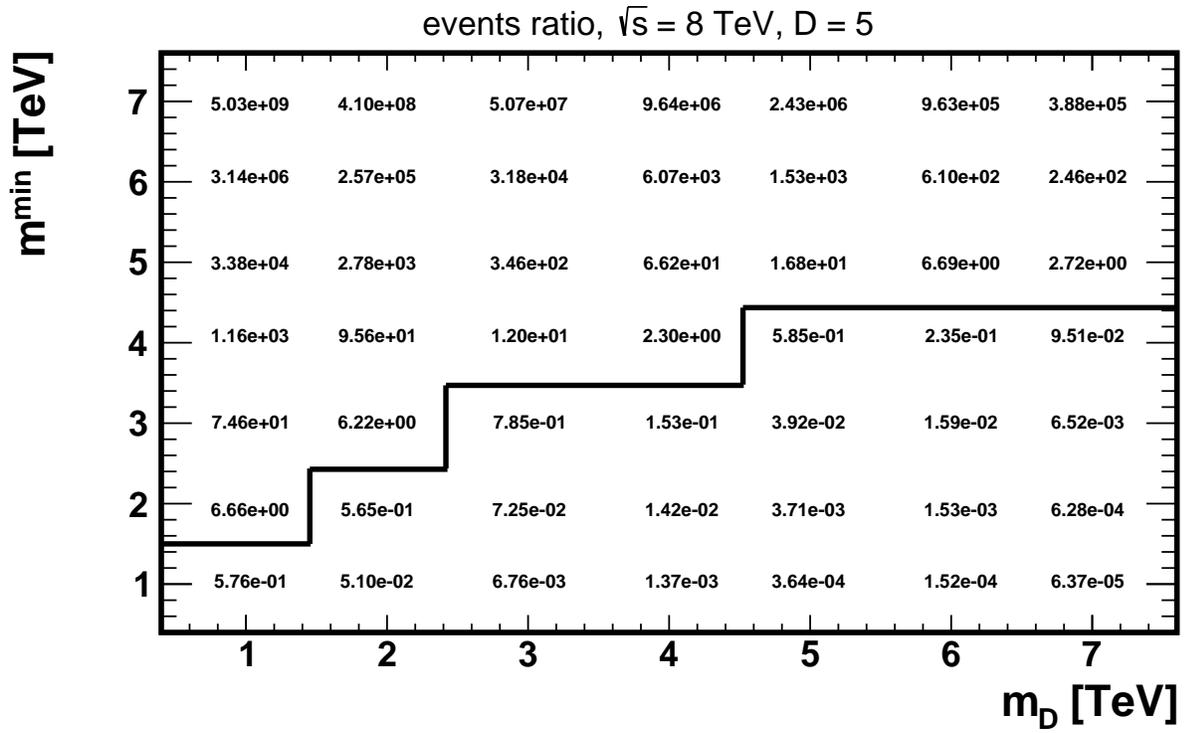}
\caption{\textit{Events ratio} as a function of $m_D$ and $m^{\rm min}$ for $D = 5$ at $\sqrt{s} = 8\,$TeV.
The thick black line indicates the cross-over.}
\label{event_ratio_dim5_8TeV}
\end{figure*}
\begin{figure*}
\centering
\includegraphics[scale=0.9]{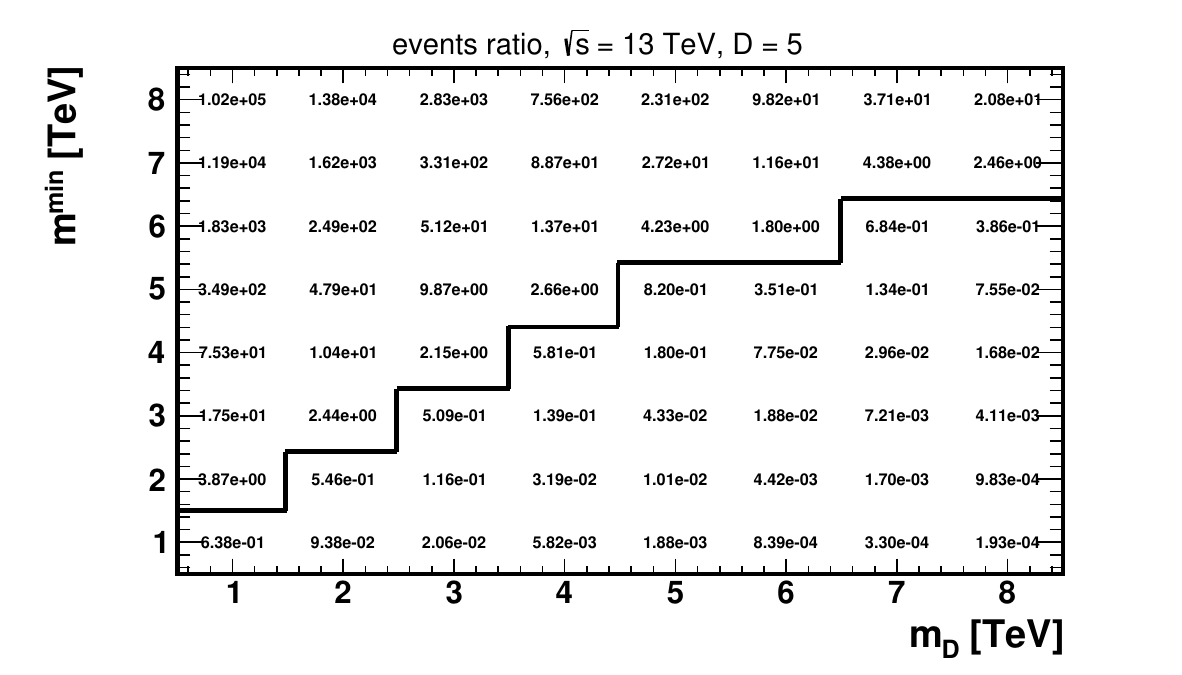}
\caption{\textit{Events ratio} as a function of $m_D$ and $m^{\rm min}$ for $D = 5$ at $\sqrt{s} = 13\,$TeV.
The thick black line indicates the cross-over.}
\label{event_ratio_dim5_13TeV}
\end{figure*}
\begin{figure*}
\centering
\includegraphics[scale=0.9]{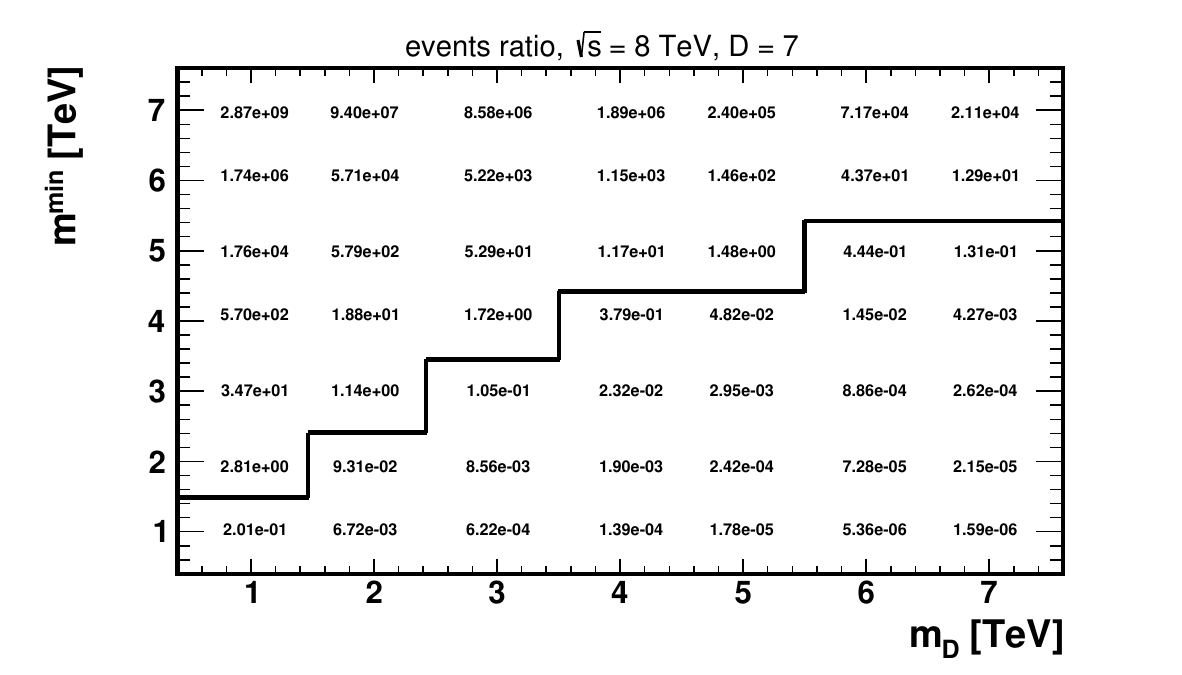}
\caption{\textit{Events ratio} as a function of $m_D$ and $m^{\rm min}$ for $D = 7$ at $\sqrt{s} = 8\,$TeV.
The thick black line indicates the cross-over.}
\label{event_ratio_dim7_8TeV}
\end{figure*}
\begin{figure*}
\centering
\includegraphics[scale=0.9]{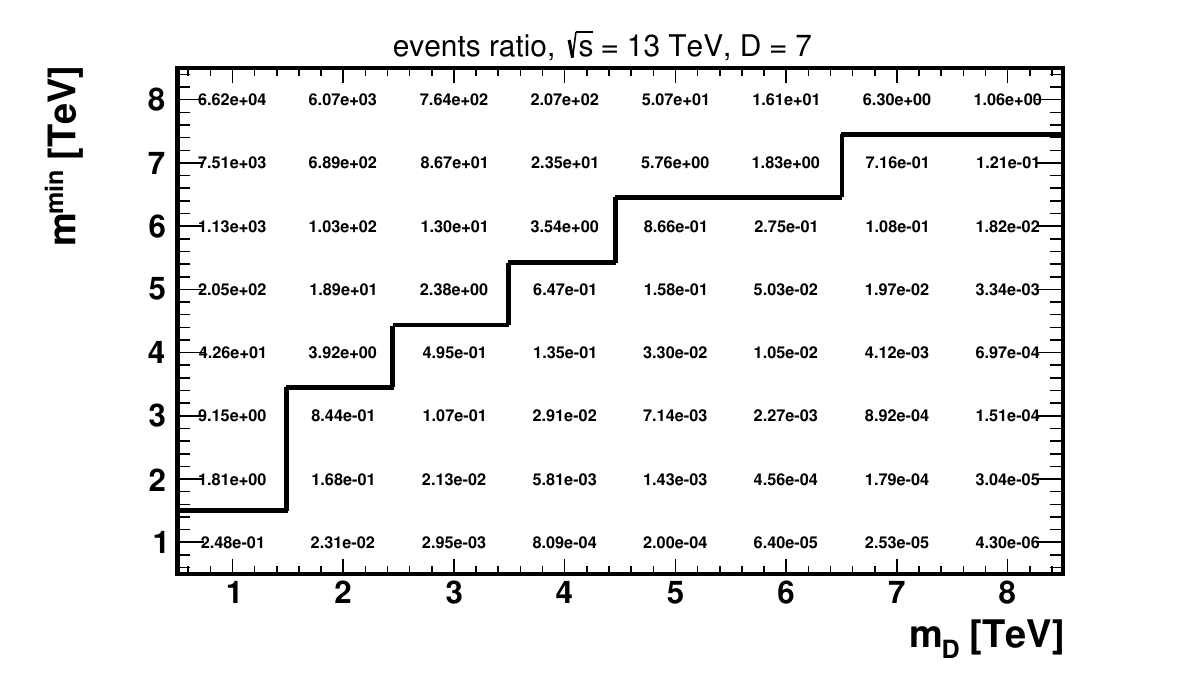}
\caption{\textit{Events ratio} as a function of $m_D$ and $m^{\rm min}$ for $D = 7$ at $\sqrt{s} = 13\,$TeV.
The thick black line indicates the cross-over.}
\label{event_ratio_dim7_13TeV}
\end{figure*}

\end{document}